Spin Filtering with Poly-T Wrapped Single Wall Carbon Nanotubes

Kazi M. Alam, Sandipan Pramanik*
Department of Electrical and Computer Engineering
University of Alberta, Edmonton, AB T6G 2V4, Canada

* Corresponding author, spramani@ualberta.ca

**Abstract**

Spin filtering is an essential operation in spintronics that allows creation and detection of spin polarized carriers. Transition metal ferromagnets are used as spin filters in most cases, though their spin filtering efficiency is only around ~50%, thereby limiting the efficiency of spintronic devices. Recently, chiral systems such as DNA have been shown to exhibit efficient spin filtering, a phenomenon often dubbed as "chirality induced spin selectivity" (CISS). In this work, we consider single wall carbon nanotubes helically wrapped with single stranded poly-T DNA. By magnetoresistance measurements we show that this system exhibits significant spin polarization of ~80%, which could be attributed to the Rashba spin-orbit interaction induced by the inversion asymmetric helical potential of the DNA. Observed spin polarization is larger than that reported before for $d(GT)_{15}$ strands. Such systems allow tailoring spin polarization by chemical means and also allow extremely localized creation and detection of spin polarization without any magnetic element and could lead to extreme miniaturization and compact integration of spintronic devices and circuits.

The central idea in the area of "spintronics" is to achieve device functionalities such as sensing, information processing, transmission and storage by utilizing spin polarizations of charge carriers[1,2]. Traditionally, inorganic materials have been used to realize spintronic devices. For example, ferromagnetic metals (such as Ni, Co, Fe and their alloys) have been used to generate (or inject) and detect spin polarized carrier populations and various non-magnetic materials (such as Cu, Si, GaAs etc.) have been used as the spin transport channel[1]. Recently, organic semiconductors (small molecules as well as long chain polymers) have been investigated heavily as potential transport channel in spintronic devices[3]. These materials have weak spin-orbit coupling and hyperfine interaction, resulting in long spin lifetime, which is an important requirement for spin based information processing[3,4]. In almost all of these experiments, transition metal ferromagnets were used as spin injectors and detectors.

However, based on recent studies, it appears that organic materials, even though they are inherently non-magnetic, can play a serious role in realizing spin injectors and detectors. It has been reported recently that chiral organic molecules, such as DNA, can perform as "spin filters", in the sense that they can generate and detect spin polarized carrier populations[5]. In particular, double stranded DNA helices have been found to offer significant spin polarization of ~60% even at room temperature[6,7]. This value is even higher than the typical spin polarization values (~50%) of transition metal ferromagnets[8]. This result is surprising since DNA molecules have small spin-orbit and hyperfine interaction, and these effects cannot be responsible for the observed large spin polarization. Significant research, both theoretical and experimental, is underway to better understand this phenomenon, which is now dubbed "chirality induced spin selectivity" or CISS. It has been surmised that the helical geometry of the molecules induces a strong inversion-asymmetric electric field, which translates to a sizable amount of Rashba spin-orbit interaction, resulting in large spin polarization[5].

Another non-magnetic system that exhibits significant spin polarization and relies on the helical field induced Rashba interaction mentioned above, is single wall carbon nanotubes (SWCNT) wrapped with single stranded DNA (ssDNA)[9,10]. In these systems carrier transport occurs via the carbon nanotube channel and the ssDNA strand induces an inversion-asymmetric helical electric field, which transforms into a Rashba spin-orbit interaction. This acts like a pseudo-magnetic field on the charge carriers and polarizes their spins. The first experimental work confirming this effect was reported by us[11], where we observed significant spin polarization (~74%) of charge carriers emanating from carbon nanotubes wrapped with d(GT)$_{15}$ strands. In this work, we further investigate this effect and address whether this effect persists with other types of ssDNA strands or when non-helical molecules are bound to the nanotube surface. As described below, our findings indicate that helical binding geometry is essential and the spin filtering effect can be further enhanced by changing the compositional details of the ssDNA strands. This offers a novel pathway to engineer spin polarization by chemical means. In addition, we show that the chiralities of the nanotubes do not play a significant role, which bodes well for large-scale device synthesis, since controlling nanotube chiralities during CVD growth is still challenging[12].

To investigate the effect of helical potential on spin polarization, we have fabricated nanotube devices, with nanotubes completely wrapped with helical ssDNA strands along their lengths. Device schematic and measurement geometry (four terminal) are shown in Figure 1a. The wrapped tubes are contacted by gold and Ni electrodes. The role of the Ni electrode is to detect spin polarization of the carriers that traverse through the wrapped tube from Au contact.

As-purchased SWCNTs (Figure 1b) are "bundled" due to highly attractive van der Waal forces and need to be separated. There exist various surfactants that can isolate bundled tubes[13,14]. Some of these surfactants are chiral[13] (such as various ss-DNA strands), and wrap the tube helically. Whereas, others (such as sodium dodecyl sulfate or, SDS, chemical formula: $(CH_3(CH_2)_{11}-SO_4Na)$ do not create any such helical binding[14]. Comparison of transport properties of both types of devices is necessary to understand the role played by the helical potential.

In our earlier study[11], we reported magnetoresistance measurements on SWCNTs wrapped by $d(GT)_{15}$ strands that consist of 30 alternating G and T bases. This study indicated creation of spin polarization as a result of $d(GT)_{15}$ wrapping. To investigate whether this effect exists for other types of ssDNA strands with different chemical compositions and whether non-helical binding with surfactants results in similar effects, in this study we use $d(T)_{30}$ (poly T) as the helical binding surfactant and sodium dodecyl sulfate (SDS) as the non-helical binding surfactant and compare the magnetotransport properties of these two types of devices.

The fabrication steps have been described in detail in our earlier report[11]. Briefly, 0.8 mg of as-purchased HiPco SWCNT bundles (Unidym Inc.) was mixed with 1ml aqueous solution, which contains ssDNA (1mg/ml $d(T)_{30}$, purchased from Integrated DNA Technologies Inc.), nuclease free water and a buffer solution (Tris, EDTA) to maintain pH of 7.5. The mixture was probe-ultrasonicated (Sonics, VC 130 PB) for 90 minutes at a power level of 8W. Next, the solution was centrifuged (Sanyo MSE Micro Centaur) for 60 minutes at 14000g in order to remove large impurities. The top part was collected for subsequent device synthesis. A drop from this solution was cast on $SiO_2$ surface, and was subsequently washed and dried to minimize the number of tubes connected between the electrodes.

Figure 1c shows a typical AFM image, indicating isolation of the tubes as a result of ssDNA wrapping. It shows an isolated nanotube, ~ 1 micron long, and helically wrapped along the entire length. This is indicated by the oblique groove pattern along the axis of the tube. This image also shows part of another isolated tube at the top left corner, and again, the same pattern is observed along the entire length of this tube. These observations are consistent with our prior work on $d(GT)_{15}$ and also with existing literature on ssDNA wrapping of SWCNTs. The ssDNA strands strongly bind with nanotube surface via (a) π-stacking between the DNA bases and the $p_z$ orbitals of the carbon atoms of the CNTs and (b) electrostatic interaction with charged sugar-phosphate backbone, resulting in helical wrapping[15,16]. Absence of any bundle in Figure 1c indicates complete wrapping of the nanotubes, because partial wrapping would have resulted in partial unbinding, which we never observe. To ensure complete wrapping and complete isolation of the nanotubes, our recipe described above used excess amount of ssDNA compared to the amounts previously reported in literature[15,16]. As expected, we have never observed any non-helical binding between ssDNA and SWCNT.

For fabrication of the final device, Ni (100nm) and Au (100nm) electrodes were sputtered on $SiO_2$(500nm)/Si wafer with ~750nm gap between them. A droplet containing wrapped carbon nanotubes (as discussed above) was subsequently cast on the gap. After 15 minutes the solution was washed out, followed by nitrogen drying and vacuum annealing (200°C for 30 minutes). Annealing results in tighter wrapping and improves electrical contact[11], as evidenced by reproducible current-voltage characteristics (**Figure S3**, Supplementary Information). Average length of the nanotubes after dispersion is ~ 500 nm, with few tubes being longer. This results in very few tubes bridging the ~ 750nm gap between the electrodes. In many cases the device ends up being open-circuit, with no tube bridging the gap. Figure 1d shows a typical FESEM image of a working device. Due to the variation in number of tubes connected, there is a variation between the current values between multiple devices. In spite of this variability, same spin filtering effect has been observed in all cases (**Figures S1, S2**, Supplementary Information).

Unwrapped and wrapped tubes have been characterized by Raman spectroscopy with 532nm excitation as shown in Figure 2. As seen in Figure 2a, radial breathing mode (RBM) frequencies have been upshifted (~ 2 $cm^{-1}$) for wrapped tubes compared to the unwrapped tubes. This upshifting of RBM modes is caused by stiffening of the tubes (and reduction in tube diameter) as a result of wrapping and has been observed before by several groups[17–20]. This shift is slightly less than what we had found in our previous study on $d(GT)_{15}$[11], which could be attributed to the difference in chemical compositions of the DNA strands. From RBM frequencies, the nominal tube diameter is estimated[21] to be ~1nm, which matches vendor's specification.

Figure 2b shows Raman *G* band feature, before and after wrapping. For carbon nanotubes, Raman *G* band is characterized by a "high frequency" $G^+$ peak at ~ 1580 $cm^{-1}$, which arises due to the vibration of C atoms along the tube axis[21]. There is also a "low frequency" $G^-$ peak that appears at a slightly lower frequency than $G^+$. The $G^-$ peak is associated with the circumferential vibrations of the carbon atoms[21]. As expected, the linewidth of the entire $G$ band has been narrowed significantly after wrapping (and isolation) due the loss of intertube interaction[19,20]. Also, the $G^-$ peak is suppressed after wrapping since the helical wrapping suppresses circumferential vibration of the carbon atoms. As estimated from Figure 2b, the relative intensity of $G^-$ to $G^+$ band for the pristine and wrapped CNTs are 0.74 and 0.57 respectively. However, this intensity reduction is smaller than the case when tubes are wrapped with $d(GT)_{15}$[11], which could be attributed to the difference in chemical compositions of the strands. The comparison of Raman RBM and $G$ bands of poly T wrapped CNTs and unwrapped CNTs confirms efficient dispersion and isolation of CNTs. This data also confirms helical wrapping, because the systematic changes described above cannot originate from some arbitrary non-helical binding.

It is important to note that the Raman $G$ band never exhibited the Breit-Wigner-Fano (BWF) lineshape, which is the signature of metallic nanotubes[17,18]. Such features are expected to appear when metallic tubes with nominal diameter of ~1nm are excited with 532nm laser. Absence of such features indicate that the tubes are predominantly semiconducting. Temperature dependent transport data described later also confirms this observation.

We also prepared another batch of devices in which nanotubes are functionalized and separated by a non-chiral surfactant - sodium dodecyl sulfate (SDS) - $CH_3(CH_2)_{11}$-$SO_4Na$. There are several aggregate morphology of SDS surrounding CNTs[14]. The most probable configurations that have been proposed in the literature are cylindrical micelles, hemispherical micelles and random layer[14]. Unlike ssDNA, none of these produces any helical binding with carbon nanotube walls. This is confirmed by AFM studies on these tubes (described later).

Figures 3(a), (c), (e), (g) show the current-voltage (*I-V*) characteristics of the Au/[poly-T wrapped SWCNT]/Ni devices in the range 16-30K. Each of these plots shows the *I-V* curves for +1.2T (solid line) and -1.2T (broken line). Magnetic field is applied in-plane of the electrical contacts. There exists a clear splitting in these characteristics, especially at low temperatures, and for a given voltage bias, device resistance is higher for +1.2T. The amount of splitting decrease as temperature is increased and at 30K and beyond, the splitting is almost negligible.

The *I-V* curves described above are highly reproducible. **Figure S3** (Supplementary Information) shows multiple *I-V* scans performed on the device described in Figure 3. The scans in **Figure S3** are overlapping each other and the variation between scans is indiscernible. Comparison of the raw data between various scans reveals that there is ~1% variation from one scan to the next, which is even smaller than the marker size used in **Figure S3**. Therefore, the *I-V* splitting observed in Figure 3 cannot be attributed to any measurement artifact.

The difference in resistance values at ±1.2T can be seen more clearly from Figures 3(b), (d), (f), (h), where device resistances (d*V*/d*I*) have been computed by numerical differentiation of the corresponding *I-V* characteristics. It is clear from these plots that the difference in resistance (d*V*/d*I*) is more pronounced in the low bias range, and gradually decreases as bias is increased. In these plots, magnitude of minimum voltage bias is 0.24V. Below this value, device current is too small to be reliably measured by our experimental setup, and hence this region has been excluded from the plot.

**Figure S4** (Supplementary Information) shows the current-voltage characteristics of the reported device (Figure 3) in small voltage range of (-0.5V – +0.5V). The difference in current for two magnetic field directions can be clearly seen. The slopes of the *I-V* curves are different at low bias values (implying existence of a resistance differential) and the curves gradually become parallel as bias voltage is increased (implying gradually vanishing resistance differential). This is consistent with the d*V*/d*I* vs *V* curves in Figures 3(b), (d), (f), (h). The range (-0.17V– +0.17V) has been excluded since current values are too small to be reliably measured in this range. As shown in **Figure S5** (Supplementary Information), typical contact resistance is ~ 2–3 orders of magnitude smaller than the actual device resistance.

Magnetoresistance in such planar devices can originate from various sources such as: (*i*) orbital magnetoresistance of the poly T-wrapped tubes, (*ii*) orbital magnetoresistance of the Au contact, (*iii*) anisotropic magnetoresistance of the Ni contact and (*iv*) any spurious Hall effect due to the fringing magnetic field lines at the vicinity of the Ni contact. To investigate the source of the splitting observed in Figure 3, we have tested a batch of Au/polyT-wrapped-SWCNT/Au control devices (Ni contact being replaced by Au). As shown in Figure 4a, no splitting in *I-V* characteristics has been observed even at the lowest measurement temperature of 16K. Multiple samples have been tested to confirm the absence of any splitting for this control device. This indicates that no orbital magnetoresistance exists in polyT-wrapped-SWCNT (at least under the measurement conditions reported here) and hence this is not responsible for the splitting observed in Figure 3. This data set also indicates that Ni contact plays an important role for the observed splitting in Figure 3.

As mentioned above, the contact resistances (**Figure S5**, Supplementary Information) of these devices are ~2–3 orders of magnitude smaller than the actual devices reported in Figure 3. Also, as expected, these contact resistances show metallic temperature dependence (not shown), which is in stark contrast with the semiconducting temperature dependence as observed in Figure 3 (also Figure 5). Clearly, the contact resistances do not play any major role in our magnetotransport measurements. Ni contacts exhibit very weak anisotropic magnetoresistance, ~1%, which is significantly weaker than ~100% magnetoresistance effect observed in our devices (described later). Thus, the mechanisms mentioned above are not responsible for the magnetoresistance effects observed in Figure 3.

Figure 4b shows *I-V* characteristics of Au/SDS-SWCNT/Ni device and no splitting has been observed even at the lowest measurement temperature of 16K. As before, multiple samples have been tested to confirm the absence of any splitting for this device. This indicates that helical wrapping with ssDNA is an important ingredient for observation of the splitting. Also, this data indicates that the observed splitting cannot arise from any spurious Hall effect due to the fringe field lines near the Ni electrode. If this were the case, the splitting should have been observed in both devices (poly T and SDS).

Figure 4c shows a typical AFM image of SDS-functionalized tubes. Height variation along the nanotube axis is observed (left nanotube), but there is no periodic helical pattern as in Figure 1c. This image also shows a second nanotube (on the right), which is completely covered by SDS, with no height modulation along the axis. These images clearly describe the non-helical and arbitrary nature of binding between nanotubes and SDS.

Figure 5 shows the zero-field *I-V* characteristics of the poly-T wrapped (Figures 5a, b) and SDS (Figure 5c) functionalized nanotubes. In both cases, *I-V* characteristics are qualitatively similar and functionalization doesn't seem to have any significant qualitative influence on zero-field charge transport. This is expected, since ssDNA and SDS are insulator in nature and these molecules are unlikely to offer a parallel path for charge transport. We also note that the channel length in our devices is ~750nm, therefore direct tunneling between the contacts via these insulating molecular chains is also unlikely. The primary charge transport channel is SWCNT, bridging both electrodes. All characteristics shown in Figure 5 exhibit semiconducting temperature dependence, which is consistent with the Raman data discussed above.

To understand the magnetoresistance effect observed in Figure 3, we have recorded *I-V* data for various magnetic field values in the range -12 to +12kG (-1.2 to 1.2 T). Figures 6a, b show device resistance (d*V*/d*I*) extracted from this data as a function of the applied magnetic field for two different bias voltages. Both scans are taken at 16K, since splitting is most pronounced at low temperatures. Figures 6 a, b, show clear *hysteretic* resistance switching around zero field and the amount of switching is large, ~100% at low temperature (16K) and low bias (0.24V). The amount of switching is larger than that reported earlier for d(GT)$_{15}$ devices[11]. The amount of switching decreases as bias is increased (as shown in Figure 6c). After switching, device resistance stays stable even if magnetic field is increased. No further switching has been observed up to field value of 12kG (Figure 6c).

As can be seen from Figures 6a, b, hysteretic resistance switching occurs at ~100G for negative field to positive field scan and at ~ −100G for positive field to negative field scan. This switching field corresponds well with the measured coercive field of Ni thin films[11]. Also, as described above, anisotropic magnetoresistance of Ni electrodes or spurious Hall effects due to fringe fields from Ni contact are not responsible for the observed effect.

The hysteretic resistance switching observed in Figure 6, is reminiscent of a typical spin valve response[22] in which one ferromagnetic layer is pinned (i.e. it does not change magnetization as the magnetic field is varied in a given range) whereas the other layer is free (i.e. its magnetization can be changed relatively easily by an external magnetic field). In absence of other mechanisms (described above) that could cause magnetoresistance switching, the only possible reason behind the observed effect is a "spin-dependent" mechanism.

It has been shown theoretically that ssDNA wrapping induces an inversion asymmetric helicoidal electric field in the channel of the nanotube, which gives rise to Rashba spin-orbit interaction in the channel[9]. This interaction acts as a pseudo-magnetic field on the charge carriers and polarizes their spins[9,10]. Spin polarization in this mechanism originates solely due to the spin-orbit interaction created by the helicoidal potential of ssDNA and the external magnetic field does not play a direct role. Therefore, spin polarization created by polyT wrapped tubes are "pinned", in the sense that they cannot be flipped by the external field. The Ni electrode, on the other hand, is "free", and its magnetization can be easily flipped by an external magnetic field. Therefore, a hysteretic magnetoresistance response, similar to that shown in Figure 6, is expected. It is to be noted that there exists an interfacial layer between the wrapped tubes and the Ni contact due to unwanted surface oxidization of the contact. This layer separates the two spin polarized materials (Ni and the wrapped tubes) and the overall device has a spin valve configuration.

Spin polarization achieved by ssDNA wrapped nanotubes can be estimated by using Julliere formula, and using Ni spin polarization value of ~33%[8]. Figure 7 shows estimated spin polarization of polyT wrapped tubes to be ~80% at low temperature (16k) and low bias (0.24V). Spin polarization value is larger than that observed earlier with $d(GT)_{15}$[11], due to stronger magnetoresistance response in the present case. Spin polarization decreases as applied bias and temperature are increased. It is to be noted that the spin polarization value estimated above (80%) is actually a worst-case estimate, because surface spin polarization of Ni should be smaller than the ideal value of 33%. In that case spin polarization of the wrapped tubes should turn out to be higher than 80%.

Based on above discussion, the wrapped tubes can be modeled as a "pinned" spin polarized material, which offers a spin-dependent barrier to the charge carriers. The spin-dependent barrier height can be estimated from Figures 3b, d, f, h by determining the bias differential to achieve a given conductance level. The bias differential turns out to be 0.12eV at the lowest bias value. Such values are too high to be explained by invoking atomic spin orbit couplings of the light atomic weight elements that comprise ssDNA-SWCNT hybrid. The experimental data should be explained by invoking spin-orbit coupling that is induced by the inversion-asymmetric helical potential of the ssDNA strands. In fact, in case of SDS functionalized tubes, which does not offer any helical wrapping, no splitting has been observed, which indicates that helical wrapping geometry is a crucial element for observation of this effect. This observation is consistent with recent magnetotransport experiments performed on DNA helices[5].

Loss of spin polarization with temperature and bias can be explained by using the spin-dependent barrier height picture described above. At higher temperatures, spin mixing between the two spin-split levels is more likely, due to phonon-assisted transitions. Enhanced spin mixing will obscure any spin-dependent barrier differential and spin selectivity property of the barrier will be lost. At higher bias, carrier energy is higher than the spin-selective barrier heights and therefore spin selectivity is absent for these carriers. However, at low bias level, carrier energy is smaller and they experience the presence of spin selective barrier, resulting in a non-zero spin polarization of transmitted carriers. This could explain why the spin polarization decreases with bias.

It is to be noted that in our experiments no effort was made to control the chiralities of the nanotubes. As described above, based on Raman and transport data, the tubes are semiconducting, but could have varying chiralities. Since we observed this effect in multiple devices (for example, Figure 3, Figure S1 in Supplementary Information), it indicates that tube chirality does not play a significant role and helical functionalization is the key ingredient. This insensitivity to tube chirality is highly beneficial, since it is extremely difficult to control tube chiralities during CVD growth process[12].

To summarize, in this work we have explored spin polarization properties of poly T wrapped SWCNT. Based on magnetoresistance measurements, we find that this inherently non-magnetic system can polarize an initially spin-unpolarized carrier population (originating from non-magnetic Au) and hence can act as a spin filter. Spin polarization of ~80% has been observed at low temperature and bias. Combining with our previous work[11] on $d(GT)_{15}$ wrapped tubes, it appears that this feature is primarily related to the helical potential that is induced by the ssDNA strands. The molecular constitution of the strands appears to have an influence on the overall effect – especially, poly T wrapping appears to generate higher spin polarization compared to $d(GT)_{15}$ wrapping. This is expected, since different chains are expected to create different electric field profile within the tube channel and result in different spin-orbit interaction strengths. Non-helical functionalization does not result in this effect. We expect to see this effect to various degrees with other DNA chains as well depending on their chemical composition and quality of wrapping. Impurity ions in the DNA solution can certainly reduce this effect, especially if these ions are from heavy atoms that can produce strong spin scattering and if these entities adhere to the tube channel during fabrication. We have carried out our experiments with high purity reagents and it is unlikely that this plays a role.

In conclusion, the main significance of the observed phenomenon is that it allows creation of spin polarized carriers without using a magnetic field, and opens the possibility of creating ultra-small spintronic devices and circuits where localized generation and detection of spin polarization could be important and which cannot be achieved by a global magnetic field. The fact that this effect is observed for two types of DNA strands (poly-T and d(GT)$_{15}$), indicates that this effect could be a general phenomenon and may manifest for a wide range of DNA strands or other helical molecules with widely varying chemical compositions. This could also be present in nanotubes made of other materials, as long as helical potential can be induced in the tube. Such combinations need to be investigated in the future to explore the possibility of obtaining even higher spin polarization and room temperature operation.

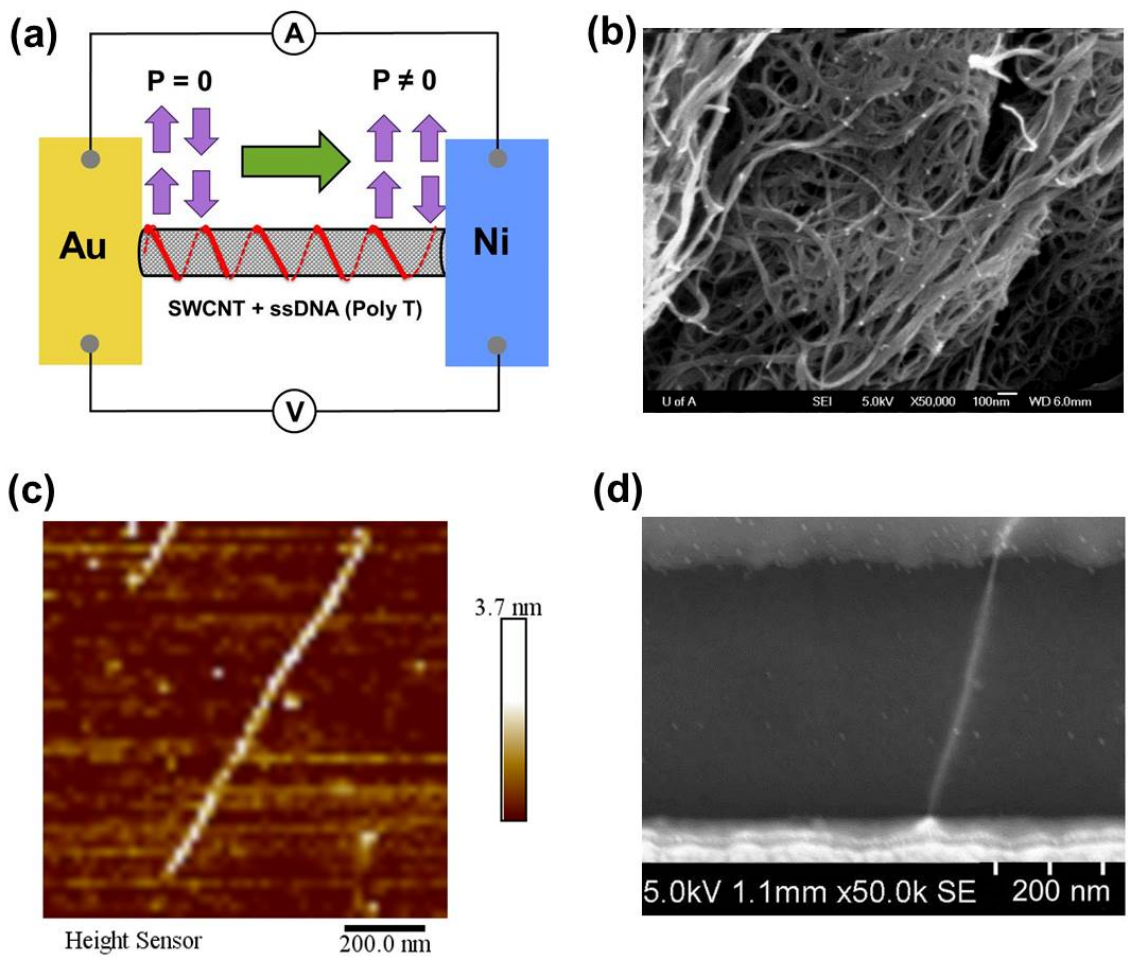

**Figure 1.** (a) Schematic depiction of the ssDNA (poly T) wrapped SWCNT spin filter device and the measurement setup. Spin unpolarized electrons are injected from Au electrode. These electrons acquire a non-zero spin polarization while traversing through the wrapped tube. The acquired spin polarization is detected by the Ni electrode, which is magnetized by an in-plane magnetic field. (b) As-purchased SWCNT bundle. (c) AFM image of separated SWCNTs, after functionalizing (wrapping) with poly T strands. Clear, periodic height modulation pattern can be observed along the entire axis of the tube, which indicates helical wrapping. Top left corner of this image shows part of a second nanotube, which also has the same pattern. (d) FESEM image of a working device, in which a poly T wrapped tube is connected between Ni and Au electrodes.

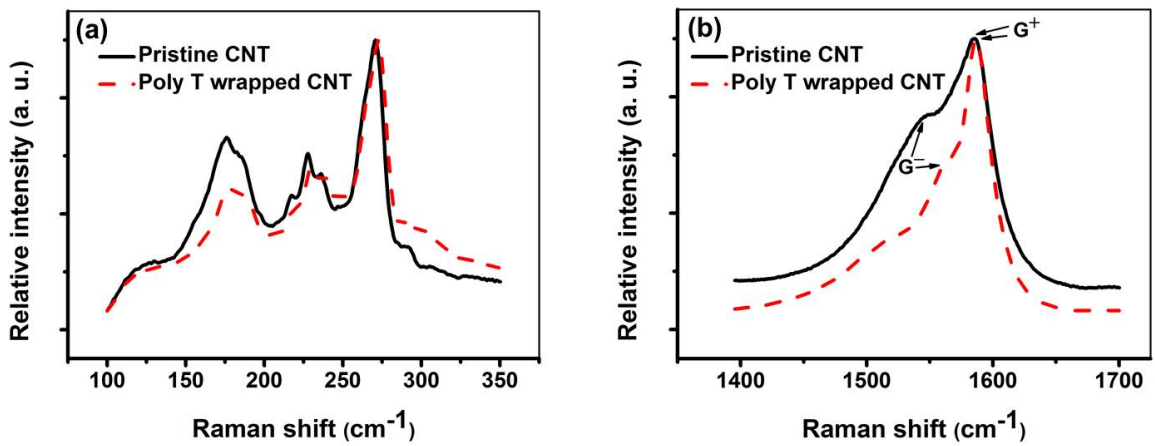

**Figure 2.** Raman spectra (532 nm) of unwrapped and wrapped nanotubes. (a) Small upshift in the RBM band frequencies has been observed as a result of wrapping. (b) *G* band linewidth narrows significantly as a result of wrapping. Also, relative intensity of $G^-$ to $G^+$ peaks reduces as a result of wrapping.

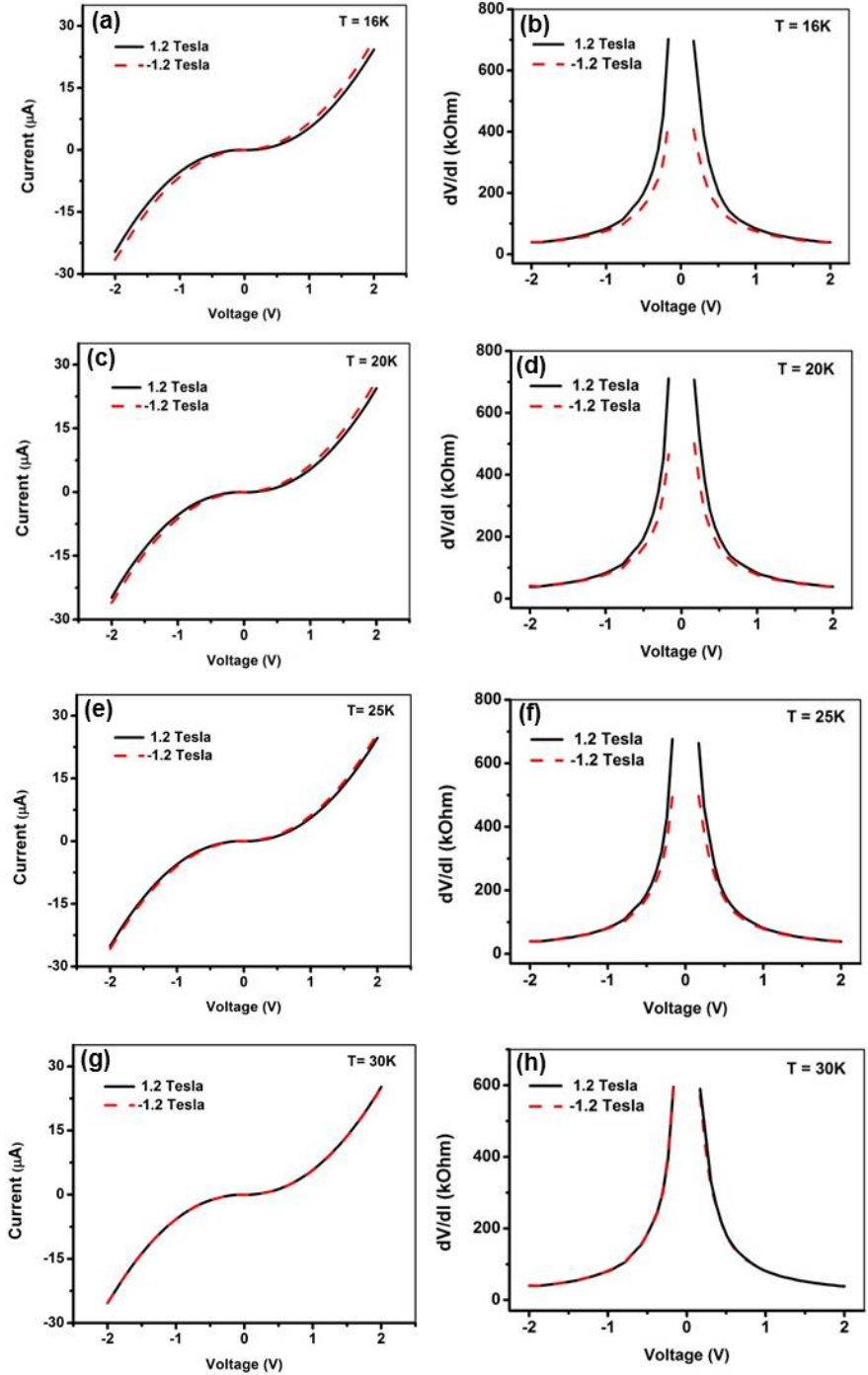

**Figure 3.** (a), (c), (e), (g) Current-voltage (*I-V*) characteristics at four different temperatures of Au/SWCNT-Poly T DNA/Ni devices. For each temperature, *I-V* characteristic depends on the applied magnetic field and there exists a split between the scans taken at +1.2 T (12 kG) and −1.2 T (−12 kG). The split decreases as temperature is increased. (b), (d), (f), (h) Resistance (computed by numerical differentiation of the *I-V* characteristics) vs bias (*V*). For a given temperature, the split is largest at lowest bias and decreases gradually with increasing bias.

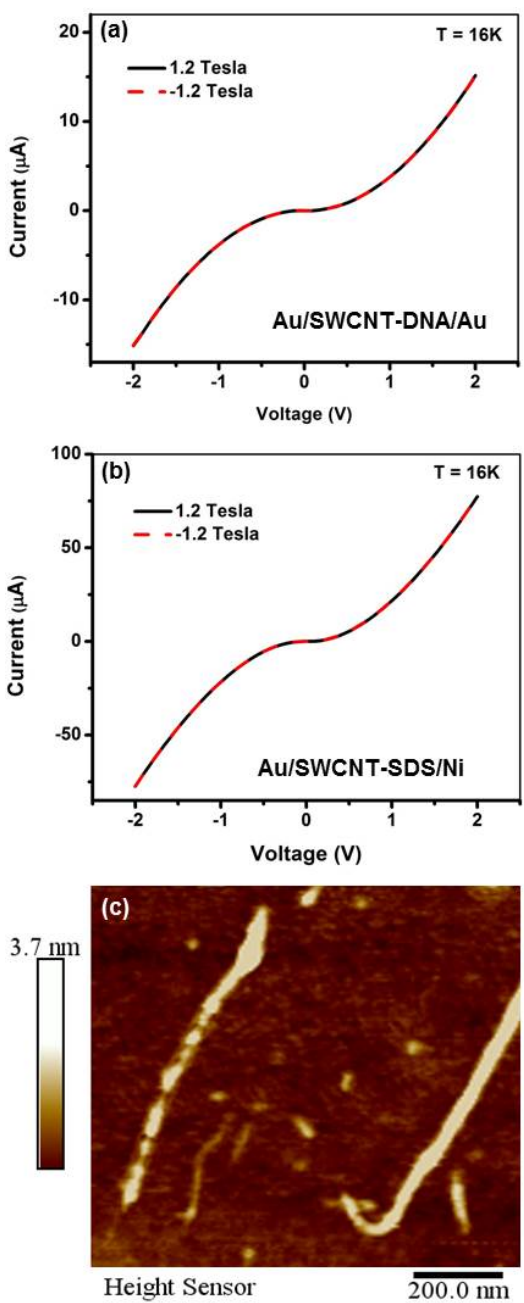

**Figure 4.** (a) *I-V* characteristics of Au/SWCNT-Poly T DNA/Au device. No splitting has been observed as a function of magnetic field even at the lowest temperature. (b) *I-V* characteristics of Au/SWCNT-SDS/Ni device. No splitting has been observed as a function of magnetic field. (c) AFM image of SDS functionalized nanotubes. Unlike Figure 1(c), no periodic height variation is present for the tube on the left. The tube on the right seems to be completely encapsulated by SDS, with no periodic height modulation. Thus, helical wrapping is absent in case of SDS functionalization.

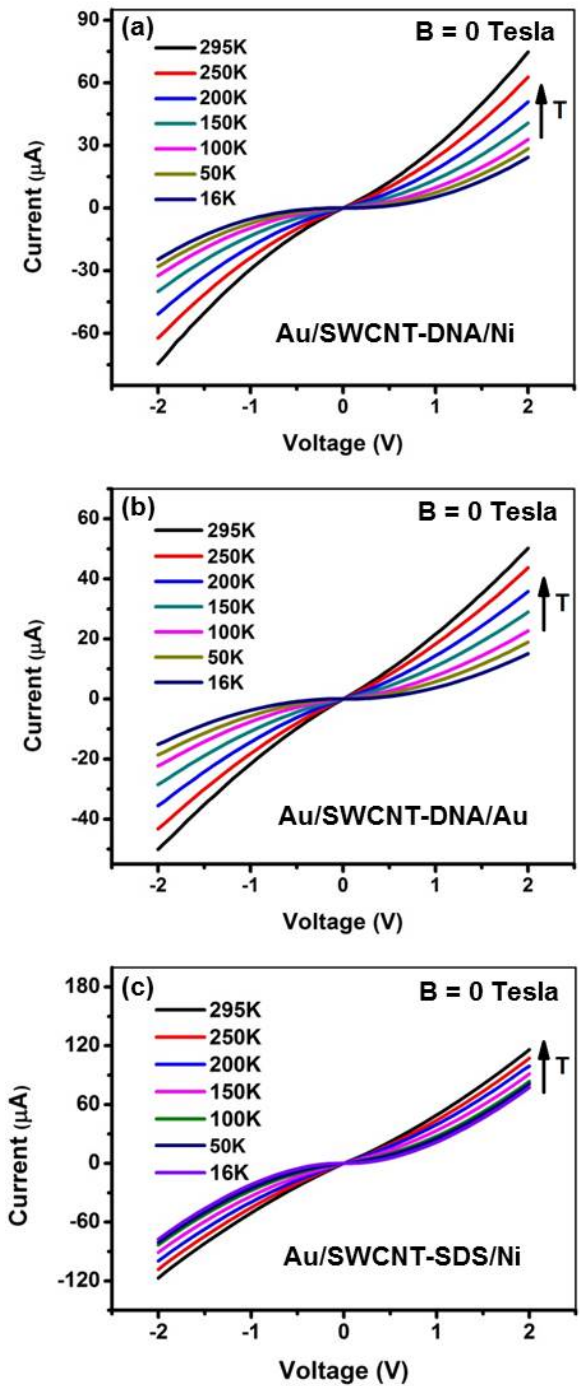

**Figure 5.** Zero-field *I-V* characteristics of (a) Au/SWCNT- Poly T DNA/Ni, (b) Au/SWCNT- Poly T DNA/Au and (c) Au/SWCNT-SDS/Ni devices. Semiconducting temperature dependence has been observed in all cases.

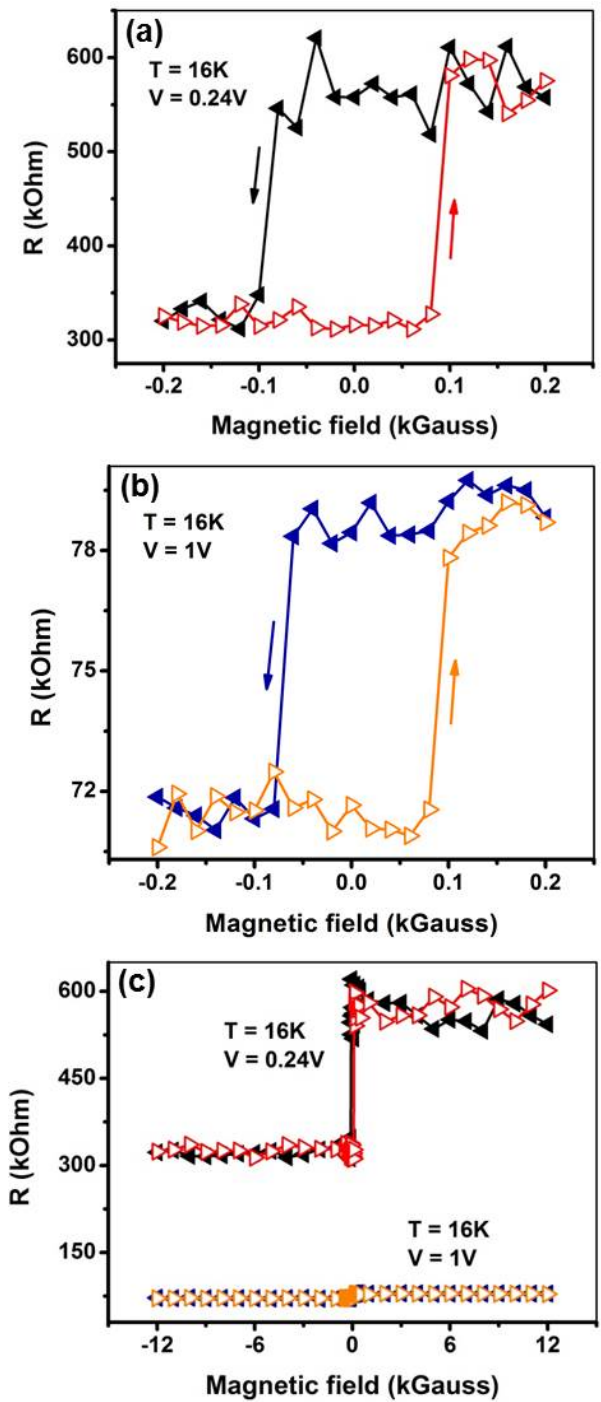

**Figure 6.** Hysteretic resistance switching in Au/SWCNT- Poly T DNA/Ni devices. (a), (b) Low field range scan at two different bias values, clearly showing the switching field. (c) Large field range scan, showing that device resistance stays stable after switching, and no further change takes place. Also, amount of switching decreases as bias is increased.

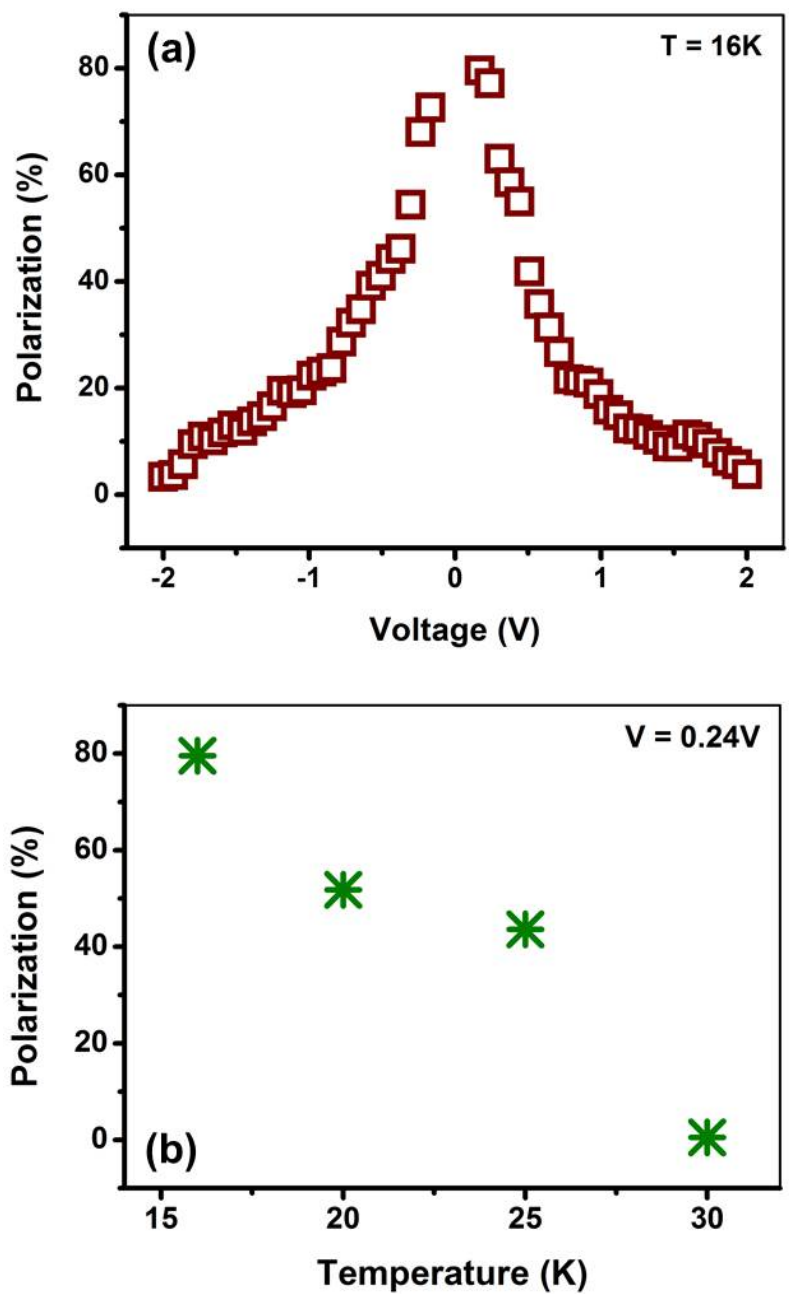

**Figure 7.** Estimated spin polarization of SWCNT-poly T composite as a function of (a) bias and (b) temperature.

# Supplementary Information.

**1. Reproducibility of the observed effect and details of *I-V* measurements.**

(a) We have measured multiple devices (~50) and have observed *I-V* splitting in ~30% of the total number of devices fabricated. **Figure S1** below shows additional data from another Au/SWCNT- Poly T DNA/Ni device. This device also shows splitting in the *I-V* characteristics as a function of magnetic field and ~ 80% spin polarization has been observed at low temperature and bias (**Figure S2**).

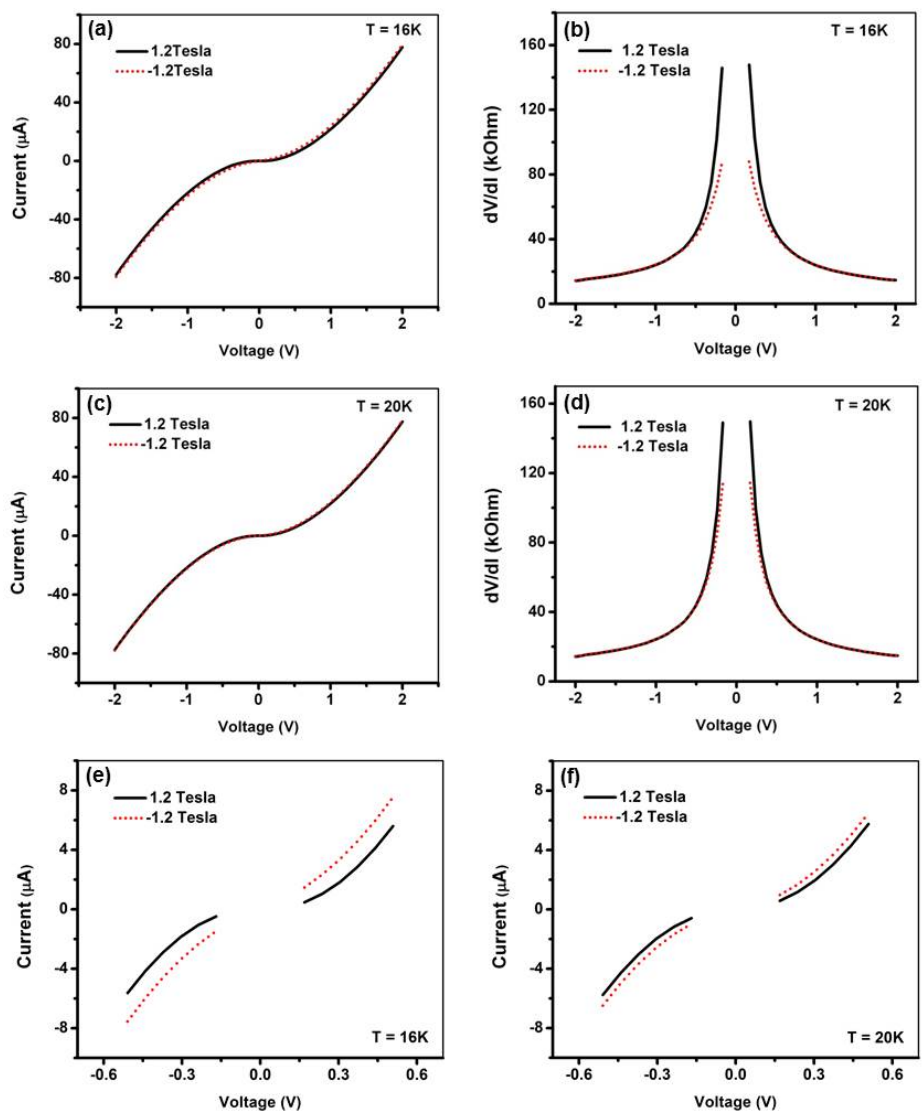

**Figure S1.** (a) – (d) *I-V* characteristics and d*V*/d*I* –*V* characteristics of a different Au/SWCNT-Poly T DNA/Ni device. (e), (f) *I-V* scans in low bias range. Current values are too low to be reliably measured near zero bias, and are hence omitted.

As discussed in the main paper, there is some variation in the current value from one device to another due to the variation in the number of tubes connected between the electrodes. We try to connect as few tubes as possible (ideally ~1), but in some cases higher number of tubes may be present. The device described in **Figure S1**, actually shows larger current than the device reported in the main paper (Figure 3). This is due to difference in the number of connecting tubes between the electrodes. Nevertheless, splitting in *I-V* curves has been observed and as shown in **Figure S2**, ~ 80% spin polarization has been found at low temperature and bias.

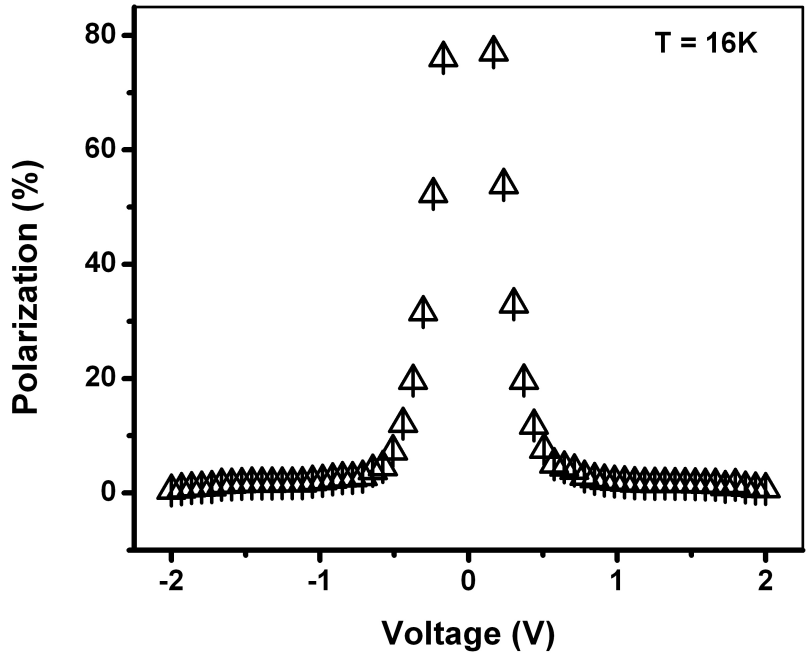

**Figure S2.** Spin polarization vs. bias voltage at 16K for the Au/SWCNT- Poly T DNA/Ni device described in **Figure S1**. Spin polarization of ~80% has been observed at low bias of 0.24V. Spin polarization decays relatively rapidly compared to the device reported in main paper. This could be due to presence of higher number of channels in this case, some of which may not remain spin selective at higher bias.

(b) For a given device, under given temperature and field value, *I-V* curves are highly reproducible. Comparison of the raw data between various scans show that there is ~1% variation from one scan to the next. This is shown in **Figure S3** below. This figure shows multiple scans of the *I-V* data reported in the main paper (Figure 3). The scans are overlapping each other and the variation between scans is indiscernible, even smaller than the marker size. Therefore, the observed *I-V* splitting cannot be attributed to any measurement artifact.

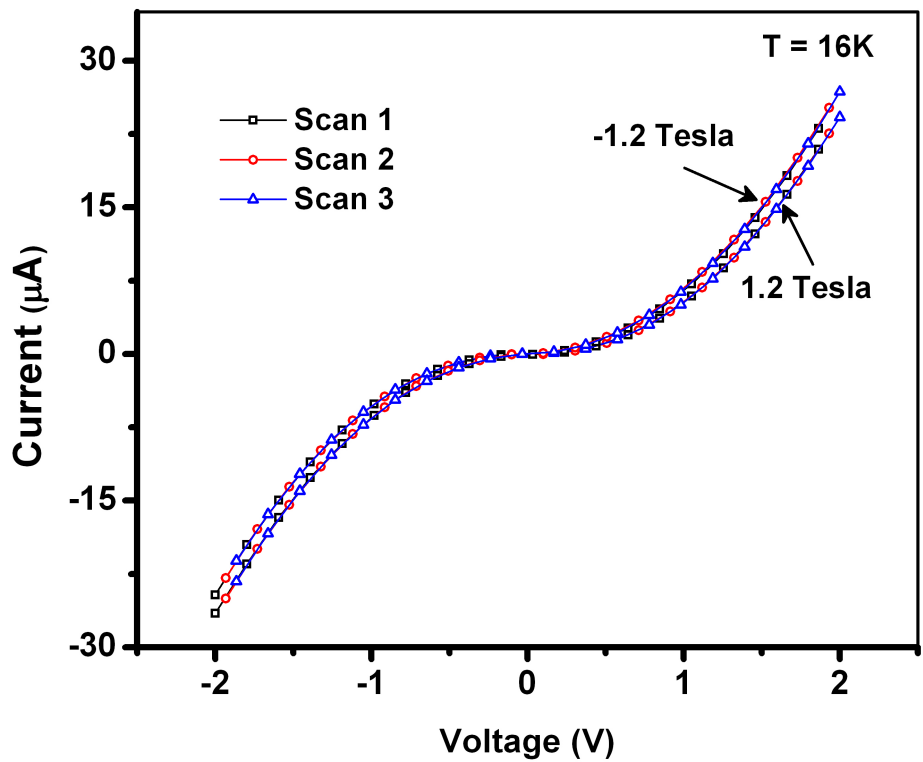

**Figure S3.** Multiple *I-V* scans of the Au/SWCNT- Poly T DNA/Ni device reported in main paper (Figure 3). The scans are overlapping each other, with ~1% difference between consecutive scans, which is smaller than the marker size used in this Figure. Thus the observed splitting cannot be attributed to any measurement artifact.

(c) **Figure S4** below shows the current-voltage characteristics of the device reported in Figure 3 (main paper) in small voltage range of (-0.5V – +0.5V). The difference in current for two magnetic field directions can be clearly seen. The slopes of the *I-V* curves are different at low bias values (implying difference in resistance) and the curves gradually become parallel as bias voltage is increased (implying gradually vanishing resistance differential). This is consistent with the d*V*/d*I* vs *V* curves in Figures 3(b), (d), (f), (h).

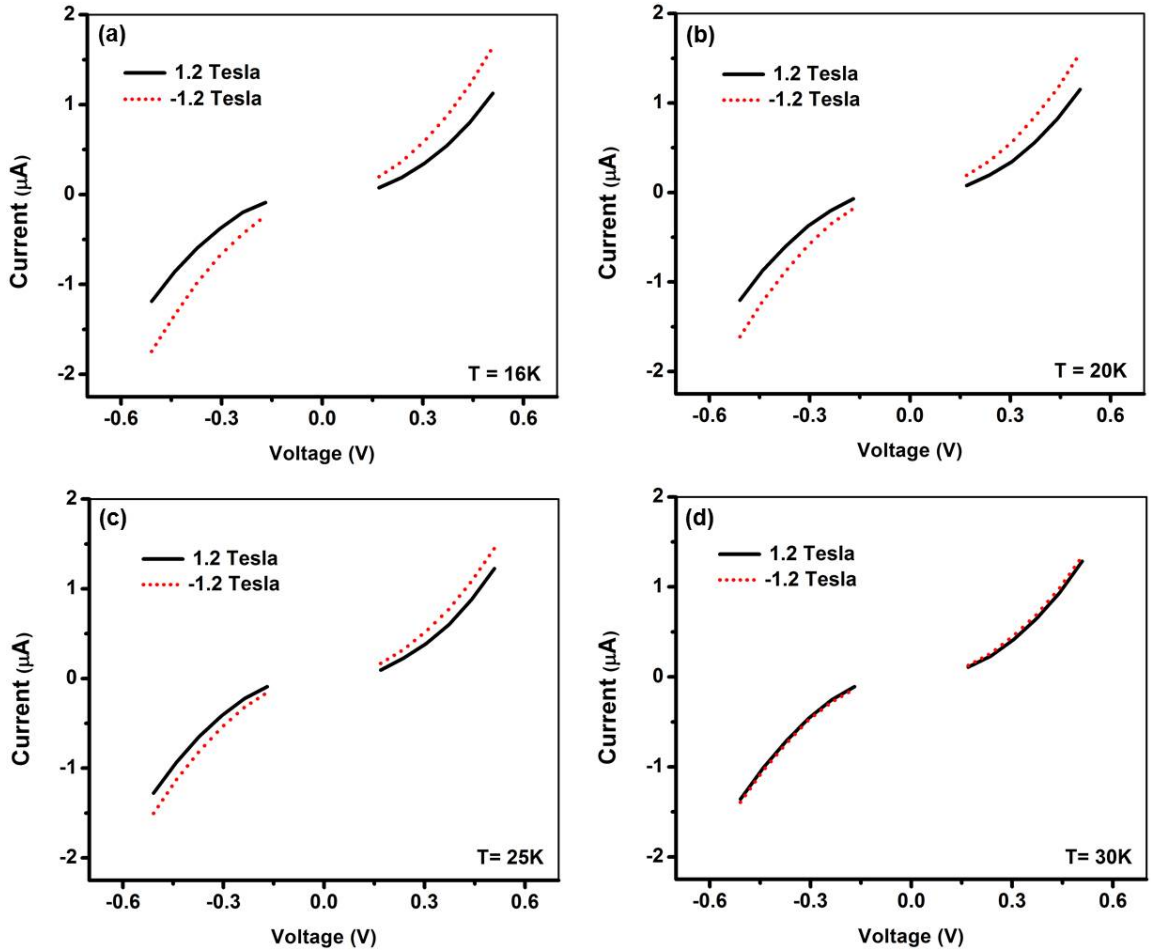

**Figure S4.** Current-voltage characteristics of the reported device (Figure 3, main paper) in small voltage range of (-0.5V – +0.5V). Current values are too low to be reliably measured near zero bias, and are hence omitted.

In **Figure S4**, we omit the region very close to zero bias, since in this range current is too small to be reliably measured by our experimental setup, as already mentioned in the main paper. This is also consistent with our d$V$/d$I$ vs $V$ curves in Figure 3 (main paper).

(d) The transport measurements are four-terminal to avoid the resistances of the metallic contacts. We have also performed two-terminal measurements on these devices. The contact resistances (of Ni and Au) extracted from this data are shown in **Figure S5**. The contact resistances are 2–3 orders of magnitude smaller than the resistance of the actual devices reported in the paper.

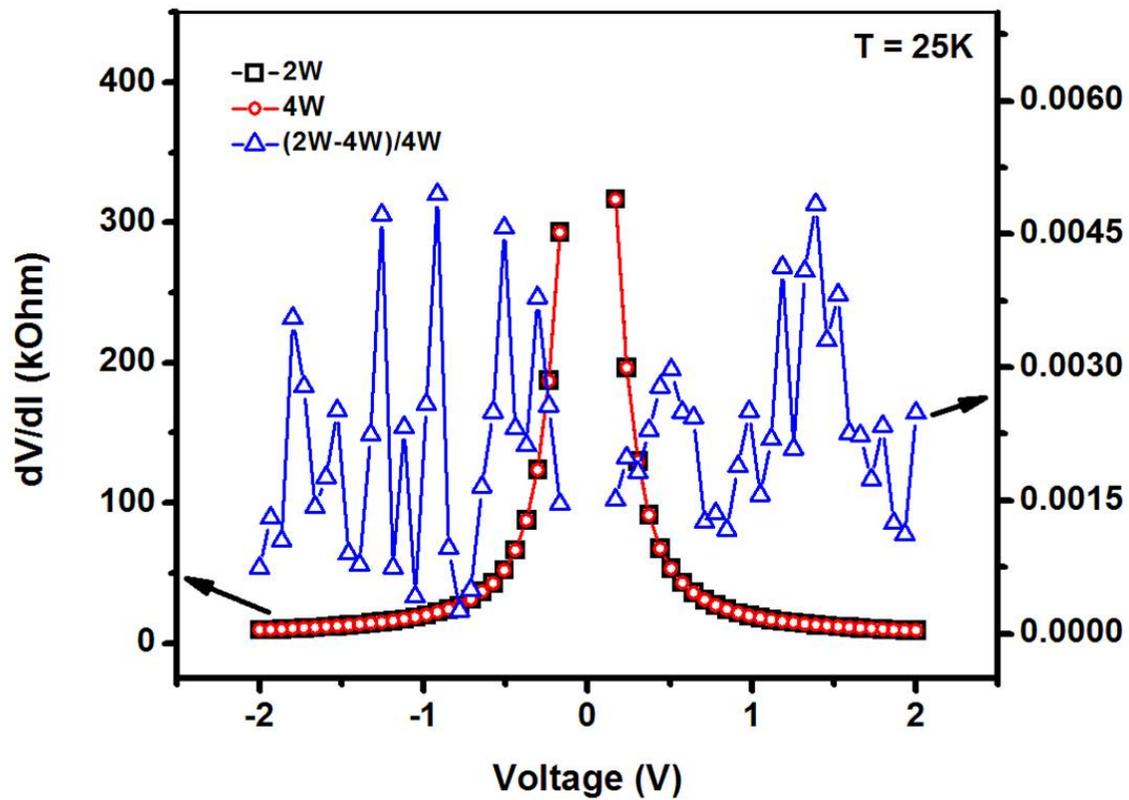

**Figure S5.** Two-wire (2W) and four-wire (4W) measurements and contact resistance normalized relative to four-wire resistance ((2W-4W)/4W). Contact resistance is 2–3 orders of magnitude smaller than the four-wire resistance.